\documentclass[doublecol]{epl2} 
\usepackage{amsmath}
\usepackage{amssymb}

\title{Fluctuation Relations at Large Scales in Three-Dimensional Hydrodynamic Turbulence}
\shorttitle{Fluctuation Relations at Large Scales in 3D HT}

\author{Alexandros Alexakis\inst{1} \and Sergio Chibbaro\inst{2} \and Guillaume Michel\inst{3}}
\shortauthor{A. Alexakis \etal}

\institute{                    
  \inst{1}Laboratoire de Physique Statistique, \'Ecole Normale Sup\'erieure, CNRS, Universit\'e P. et M. Curie, Universit\'e Paris Diderot, Paris, France\\
  \inst{2} Université Paris-Saclay, CNRS,
UMR 9015, LISN, F-91405 Orsay Cedex, France\\
  \inst{3} Sorbonne Université, CNRS, Institut Jean Le Rond d'Alembert, F-75005 Paris, France
}

\abstract{
It has long been conjectured that, in three dimensional turbulence, velocity modes at scales larger than the forcing scale follow equilibrium dynamics. Recent numerical and experimental evidence show that such modes share the same mean energy and therefore support this claim, but equilibrium dynamics does not reduce to equipartition of energy. In this work, a large set of direct numerical simulations is carried out to investigate if fluctuation-dissipation relations and the fluctuation theorem also apply at these scales. These two results link out-of-equilibrium properties of a forced system with its behavior at equilibrium. Both relations are verified quantitatively by the results of our simulations, 
further supporting that large scale modes display equilibrium dynamics.
They provide new tools to characterize both the mean value and the fluctuations of the injected energy by a large scale force acting on turbulence driven by small scale random noise.}

\begin{document}

\maketitle

\section{Introduction}

Hydrodynamic turbulence is one of the most fundamental problem in macroscopic physics~\cite{Frisch_1995}, crucial in geophysics and engineering~\cite{pope2000turbulent,vallis2017atmospheric}. 
It has been extensively studied both experimentally and numerically,
unveiling the direct cascade, that is the transfer of energy from the forcing scale at which energy is injected to the small scales where it is dissipated~\cite{kolmogorov1941dissipation,monin2013statistical,alexakis2018cascades,Verma_ET}.
Scales larger than the forcing one, thereafter referred to as large scales, are found essential in some configurations, notably in geophysics and astrophysics~\cite{moffatt1978magnetic,pedlosky2013geophysical}. Being outside of this direct cascade, these large scales are subject to neither forcing nor significant viscous damping. It is therefore tempting to model their dynamics in the framework of the Truncated Euler Equations (TEE), a Garlekin truncation of the incompressible Euler equation describing a finite number of modes that interact nonlinearly with strictly zero dissipation and external force \cite{Frisch_1995}. Direct numerical simulations of the TEE exhibit some expected features of large scale turbulent flows: smooth initial conditions evolve as a result of nonlinear interactions to spread energy over all of the available modes \cite{Cichowlas2005,Murugan2023}, eventually reaching a chaotic steady-state characterized by its thermodynamic temperature \cite{Murugan2021}. The accuracy of this modeling remains an open question \cite{rose1978fully,mccomb2014homogeneous,thalabard2015statistical, Ding_2023} but it would provide an entire new set of predictions regarding turbulence, since, contrary to the Navier-Stokes equations, it can be treated using equilibrium statistical mechanics~\cite{lee1952some,hopf1952statistical,kraichnan1973helical,bouchet2012statistical}.

%\revision{add text here}

Recent numerical and experimental investigations report equipartition of energy at large scales~\cite{Dallas_2015,cameron2017effect,Alexakis_2020, Gorce_2022}, as predicted by the TEE, at least provided some constraints on the forcing mechanism are put \cite{Alexakis_2019,Hosking_2023}. TEE have also been found able to capture bifurcation between turbulent states \cite{Shukla2016Statistical,dallas2020transitions,vanKan2022geometric}. Yet, it remains to assess to which extent linear-response theory may be applied~\cite{kraichnan1959classical,kraichnan1959structure, marconi2008fluctuation}: the first goal of the present work is to investigate the fluctuation-dissipation theorem for large scales in turbulence.

Whereas equipartition of energy is restricted to system at thermal equilibrium and linear response to systems close to equilibrium, the derivation of the Fluctuation Relations (FR) has shown that statistical mechanics also provides constraints on time-reversible systems arbitrarily far from equilibrium~\cite{evans1993probability,gallavotti1995dynamical,jarzynski1997nonequilibrium,kurchan1998fluctuation,lebowitz1999gallavotti,chetrite2008fluctuation}. 
. They successfully capture the statistics of the 
entropy production of a broad range of out of equilibrium systems, such as electrical circuits \cite{Ganier_2005}, colloidal particles \cite{Wang_2002, Carberry_2004} or molecular motors \cite{Hayashi_2010}. 
However, their practical use is limited to systems whose typical energy is of the order of $k_b T$, with $k_b$ the Boltzmann constant, and $T$ an equilibrium temperature. In standard conditions that means that only specific systems, most notably microscopic ones, may be adapted to test such relations without introducing ad-hoc or phenomenological parameters. This discards most macroscopic systems for which dynamics is irreversible~\cite{ciliberto1998experimental,falcon2008fluctuations,Barrat_2008,bandi2009probability,shang2005test,xu2014,zonta2016entropy,peinke2019fokker} but may apply to large scales in turbulence: the second goal of the present letter is to test a FR in this context.
 
More specifically, we examine
both the Fluctuation Dissipation Relation (FDR) and the Transient Fluctuation Theorem (TFT)~\cite{Evans_2002}. The FDR  links equilibrium correlations to the response of the system to an external perturbation~\cite{kubo1966fluctuation}. The more general TFT characterizes the probability of the injected power being negative when integrated over a finite duration.  These relations are tested through direct numerical simulation of fully turbulent flows governed by the hyperviscous NSE.

\section{Methods}

We investigate a turbulent flow described by the 
velocity field $\mathbf{u}(x,y,z,t)$ evolving in triple periodic domain $ \in [0, 2\pi]^3$ following the hyper-viscous and incompressible NSE:
\begin{align}
&\frac{\partial \mathbf{u}}{\partial t} + (\mathbf{u} \cdot \nabla) \mathbf{u} = 
{\bf F}
%\mathbf{F}_\mathrm{ls} + \mathbf{F}_\mathrm{ss}
- \nabla P - 
%\frac{1}{Re}
\nu_2 \nabla^4 \mathbf{u}, \\
&\nabla \cdot \mathbf{u}=0.
\end{align}
Here $P$ is the pressure, $\nu_2$ is the hyper-viscosity
and $\bf F$ is the forcing. 
The flow is simulated using the pseudospectral parallel code {\sc ghost} \cite{mininni2011hybrid} with a second order Runge-Kutta method. Because we are interested in very long statistics we have limited our resolution to moderate values of grid resolution $N$, being set to either 128 or 256. For this reason, the use of hyper-viscosity is mandatory to display a developed turbulence regime
and to guarantee that viscous effects play no role at large scales.
In order to decompose the flow among different degrees of freedom, we use 
the Fourier representation of $\mathbf{u}(\mathbf{x},t)$ 
\begin{align}
\mathbf{u}(\mathbf{x},t) &= \sum_{\mathbf{k} \in\mathbb{N}^{3}} 
%\left[ 
\mathbf{\hat{u}}(\mathbf{k},t)e^{i \mathbf{k}\mathbf{x}} .
%+ \mathbf{\hat{u}}(-\mathbf{k},t) e^{-i \mathbf{k}\mathbf{x}} \right]
%\\
%& =\sqrt{2} \sum_{\mathbf{k} \in\mathbb{N}^{3}} \left[R(\mathbf{k},t) \cos( \mathbf{k}\mathbf{x}) + I(\mathbf{k},t) \sin( \mathbf{k}\mathbf{x}) \right],
\end{align}

\begin{figure}[t]
    \begin{center} \includegraphics[width=.48\textwidth]{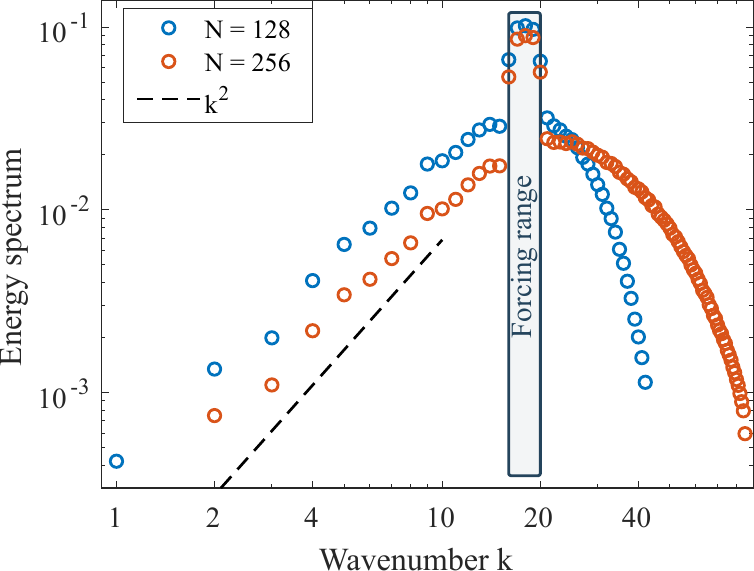} \end{center}
    \caption{Mean energy spectrum of the base state.}
   \label{Fig_PSD}
 \end{figure}

The forcing is split into two parts ${\bf F= F}_\mathrm{ss} + {\bf F}_\mathrm{ls}$. ${\bf F}_\mathrm{ss}$ acts at large wavenumbers $\bf k$ (small scales) 
such that $k_f \le |{\bf k}| <k_f+\delta k_f $ (with $k_f=18$, and $\delta k_f=2$).
It is random and delta-correlated in time, and provides a constant energy injection rate that drives the direct cascade. $\mathbf{F}_\mathrm{ls}$ on the other hand is a time independent Kolmogorov forcing, of functional form given by ${\bf F_\mathrm{ls} }= \sqrt{2} F_0 \cos(k_{0} y){\bf e}_x$ where ${\bf e}_x$ is the unit vector in the $x$ direction, $F_0$ is an amplitude and $k_{0}$ a chosen wavenumber.

The large scale force $\mathbf{F}_\mathrm{ls}$ is first set to zero and the turbulent velocity field is only driven by the small scale force $\mathbf{F}_\mathrm{ss}$. The resulting statistically stationary steady state is thereafter referred to as the base state, and its time-averaged energy spectrum is reported in Fig. \ref{Fig_PSD}. As previously observed, it shows that large scale modes approximately share the same mean energy (the $k^2$ scaling law). Small deviations from this law are attributed to the finite width of the forcing range and the limited scale separation \cite{Alexakis_2019}. In what follows we monitor the amplitude $R(t)$ of the mode subsequently forced by ${\bf F}_\mathrm{ls}$,  defined as
\begin{align}
R(t) =\sqrt{2} \Re[{\bf e}_x \cdot  \mathbf{\hat{u}}(k_0\mathbf{e}_y,t)  ],
\end{align}
where $\Re$ stands for the real part, and $\mathbf{e}_y$ is the unit vector in the $y$ direction. The auto-correlation functions $g(\tau) = \langle R(t) R(t+\tau)\rangle$ is computed, with $\langle \cdot \rangle$ a time-average,  as well as the thermodynamic temperature $T = g(0)$ and the integral $G \equiv \int_0^{\infty} g(t) \mathrm{d}t$ estimated as
\begin{equation}
G = \int_0^{\tau_\mathrm{max}} g(t) \mathrm{d}t, 
 \end{equation}
with $\tau_\mathrm{max}$ a finite duration over which $g$ is converged.

\section{Fluctuation Dissipation Relation} 

\begin{figure}[t]
    \begin{center} \includegraphics[width=.48\textwidth]{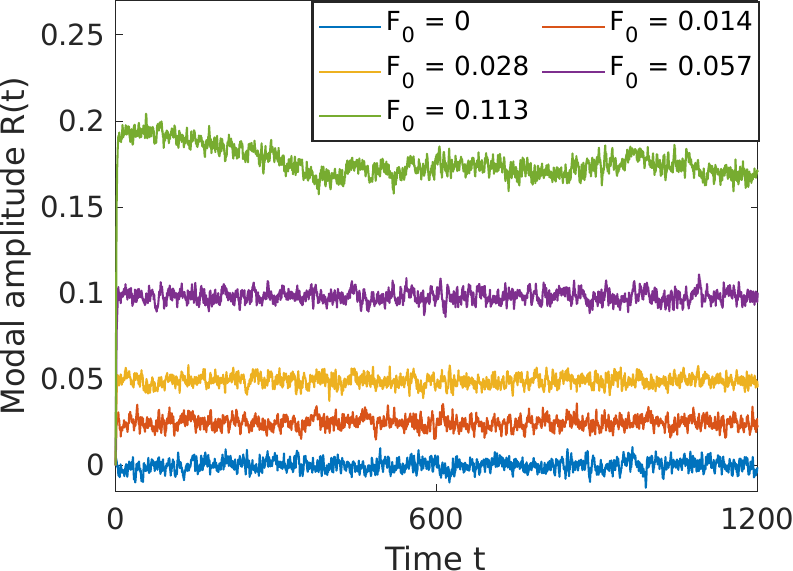} \end{center}
    \caption{Time series of $R(t)$ once the large-scale force of amplitude $F_0$ is turned on ($k_0 = 4$). Note that $F_0 =0$ corresponds to base state (no large scale force), based on which are computed both the temperature $T$ and the integral $G$.}
   \label{Fig_GK_tseries}
 \end{figure}

The nonequilibrium response of the mode of amplitude $R(t)$ is now investigated. Once the base state reached and both $T$ and $G$ computed, the large scale force $\mathbf{F}_\mathrm{ls}$ is turned on and acts on this mode only: the energy of this mode $R(t)^2/2$ is now subject to an additional injected power $F_0 R(t)$. This mode eventually reaches a statistically steady state of non-zero mean drift $\langle R\rangle $, as evidenced by some time series reported in Fig. \ref{Fig_GK_tseries}. 
 \begin{figure}[htb]
    \begin{center} \includegraphics[width=.49\textwidth]{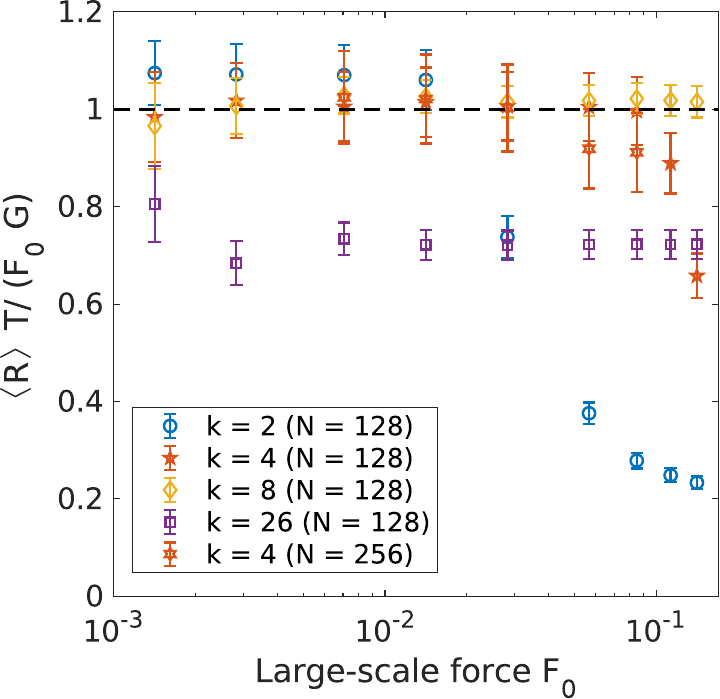} \end{center}
    \caption{Validity the Fluctuation Dissipation Relation for various wavenumbers $k_0$, amplitudes of large scale force $F_0$ and grid resolutions $N$. }
   \label{Fig_GK_comparison}
 \end{figure}
The FDR predicts that, in the limit $F_0 \rightarrow  0$, this quantity evolves according to~\cite{Kubo_1957}
\begin{equation}
\langle R \rangle = \frac{G }{T } F_0,
\label{eq:GK}
\end{equation}
i.e., that the mean injected power is $\langle F_0 R \rangle = G F_0^2/T$.  To test this relation we performed several DNS varying the amplitude  of the large scale force $F_0$,  the forcing wavenumber $k_0$ and the spatial resolutions $N$. In figure \ref{Fig_GK_comparison}, we plot the ratio $\langle R\rangle T/F_0G$ that based on the relation (\ref{eq:GK}) should be unity.
Results show an excellent agreement with the FDR in the limit $F_0 \rightarrow  0$, except for $k_0 = 26$.
This mode is in fact in the direct cascade ($k_0>k_f$), so that there is a marked departure from equipartition and detailed balance. Yet, a linear relation appears still to hold though with a non-unity coefficient, because of the sweeping effect~\cite{kraichnan1959structure,kraichnan2000deviations}. This observation is in agreement with previous numerical investigations of the linear response in the direct energy cascade using shell models \cite{Matsumoto_2014,Cocciaglia_2023} and the incompressible Navier-Stokes equations \cite{Matsumoto_2021}: in both cases, the FDR is found not to hold quantitatively but provides the right order of magnitude.

A departure from the linear relation (\ref{eq:GK}) is also observed for large forcing amplitudes $F_0$. This is expected, since eq. (\ref{eq:GK}) is valid in the
$F_0\to 0$ limit.
We have verified that in correspondence with these perturbation amplitudes (e.g., $F_0\gtrsim 0.1$ for $k_0=4$) there is a nonlinear energy transfer to other modes, such that the detailed balance is broken.

\section{Transient Fluctuation Theorem}
 
\begin{figure}[t]
    \begin{center} 
\includegraphics[width=.48\textwidth]{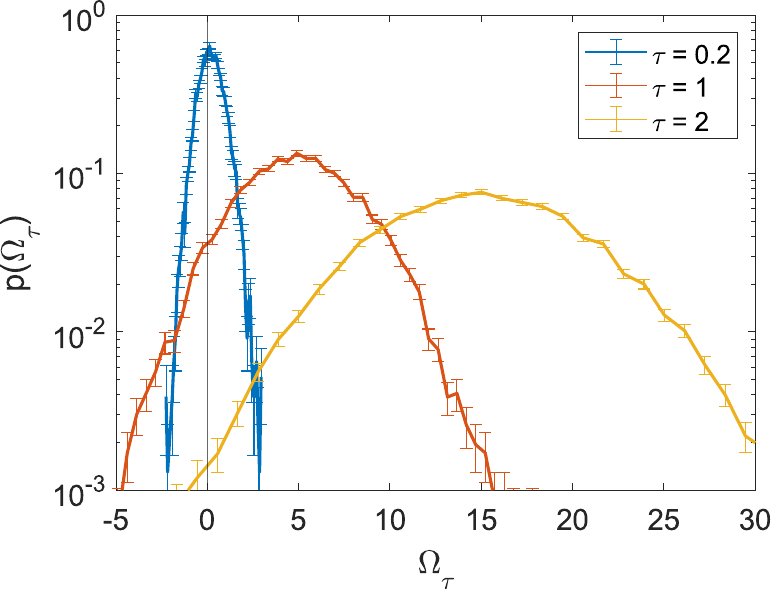}
    \caption{PDFs of the normalized work $\Omega_\tau$ of the large-scale force  for various time horizons $\tau$.}
    \label{Fig_TF_PDF}
   \end{center}
\end{figure}

The wavenumber $k_0$ is now set to $8$ and the statistics of the entropy production (or dissipation function) integrated over the interval $t\in [0,\tau]$ is investigated. This quantity reduces in this system to the normalized work done by this large scale force over the interval $t\in [0,\tau]$,
\begin{equation}
\Omega_\tau = \frac{\int_0^\tau  F_0 \times R(t) ~\mathrm{d}t}{T}.
\end{equation}
Its specific value depends on the realization, i.e., on the microstate at time $t=0$ when the large scale force is turned on. In figure \ref{Fig_TF_PDF} we plot $p(\Omega_\tau)$, the Probability Distribution Function (PDF) of $\Omega_\tau$, for various time horizons $\tau \in [0.2,1,2]$ similar to the autocorrelation time of this mode at equilibrium ($G/g(0)=0.8$).  
To calculate these PDFs we used an ensemble of over 9000 different realizations 
that were varying in initial conditions.
The different initial conditions used were  obtained from a long run with ${\bf F}_\mathrm{ls}=0$
sampled every $\tau_{sample}= 2/(u_{rms}k_0) $. 

\begin{figure}[ht]
    \begin{center} 
\includegraphics[width=.48\textwidth]{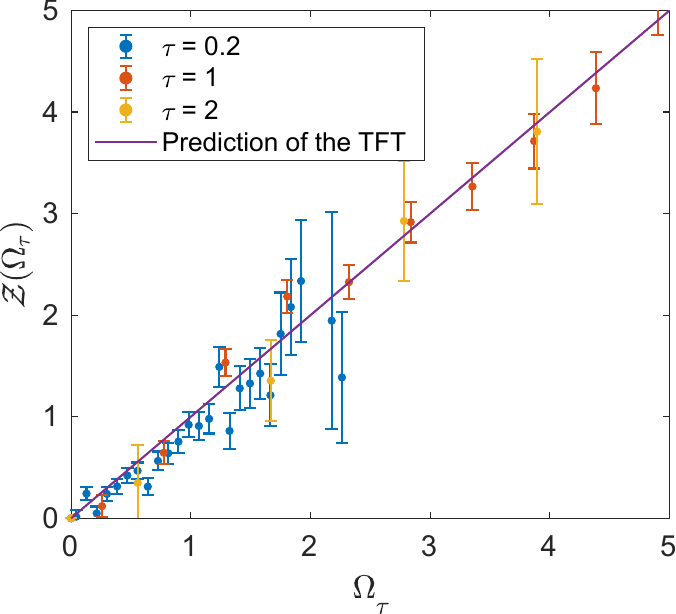}
    \caption{Measure of the asymmetry of $p(\Omega_\tau)$ and comparison to the prediction of the TFT (no fit parameter).}
    \label{Fig_TF_comparison}
   \end{center}
\end{figure}

Both the mean value of $\Omega_\tau$ and its variance increase with $\tau$, 
 such that negative values of $\Omega_\tau$ are observed even for large $\tau$.
 However, the fraction of realizations such that $\Omega_\tau<0$ decreases as $\tau$ increases, going from $38\%$ for $\tau = 0.2$ to $0.3\%$ for $\tau=2$. These events correspond to rare realizations for which
 the large scale force actually removes energy from the system.
The behavior of $p(\Omega_\tau)$ is constrained by a fluctuation theorem.
Different forms of the fluctuation theorem can be derived \cite{marconi2008fluctuation}, the most relevant being in this context the Transient Fluctuation Theorem, according to which the asymmetry measure $\mathcal{Z}(\Omega_\tau)$ given by
\begin{equation} 
\mathcal{Z}(\Omega_\tau   )\equiv
\mathrm{ln}\left(
\frac{p(\Omega_\tau)}{p(-\Omega_\tau)} 
\right)
\end{equation}
is such that
\begin{equation} 
 \mathcal{Z}(\Omega_\tau  )=\Omega_\tau.
\label{TF_prediction}
\end{equation}
This relation gives a measure of the asymmetry of the PDF of $\Omega_\tau$ between positive 
and negative values, implying that negative values $- \vert \Omega_\tau \vert $ 
are exponentially less likely to occur than positive ones $\vert \Omega_\tau \vert$.
We test this relation by calculating  $\mathcal{Z}(\Omega_\tau)$ for different values
of $\Omega_\tau$ and different integration times $\tau$
using the large ensemble formerly described.
The comparison of DNS results with the prediction given by eq. (\ref{TF_prediction}) is excellent, as shown in Fig. \ref{Fig_TF_comparison}. 
Note that it involves no fit parameter. We further remark that although the TFT is derived for an arbitrary value of $\tau$, it becomes increasingly difficult to test as $\tau \rightarrow \infty$, since the number of events such that $\Omega_\tau <0$ vanishes~\cite{zonta2016entropy}. Present simulations are able yet to verify 
the relation up to $\Omega_\tau=5$, something reached only on simple models~\cite{gallavotti2014equivalence}.

\section{Conclusions}

Both the Fluctuation-Dissipation relation and the Transient Fluctuation Theorem have been found to hold for the large scales of 3D hydrodynamic turbulence. 
Although turbulent states are very far from equilibrium and undergo an irreversible dynamics, our results show that scales larger than the one at which energy is injected are described with negligible errors as truncated Euler modes. Then, they can be accurately analysed through equilibrium statistical mechanics, since detailed balance applies.
This picture was previously conjectured from energy spectra showing an approximate equipartition of energy; the results obtained for finer details of the dynamics investigated here demonstrate it unambiguously.  
These finding are also  important from a practical perspective since they predict the mean value and statistics of the power injected by a large scale force in turbulence. It could be experimentally investigated by adding a large scale propeller in the experiment in which large scale equipartition was observed \cite{Gorce_2022}.  
Moreover, since the proper variables of the system are identified, our work paves the way to use equilibrium statistical mechanics and maximum entropy techniques to model geophysical large scales, as already successfully done for the simpler 2D case~\cite{bouchet2012statistical}.

\acknowledgments
AA was supported by the Agence nationale de la recherche project LASCATURB
No. ANR-23-CE30.

\end{document}